\documentclass[conference]{IEEEtran} 
\IEEEoverridecommandlockouts
\usepackage{relsize}
\usepackage{caption}
\usepackage{subcaption}
\usepackage{graphicx}
\usepackage{amssymb}
\usepackage{mathrsfs}
\usepackage{mathtools, nccmath}
\usepackage[hidelinks]{hyperref}

\usepackage{setspace}

\usepackage{amsmath,mathtools}
\usepackage{widetext}
\usepackage[usenames,dvipsnames]{color} 
\usepackage{ifthen}

\usepackage{tabularx}
\usepackage{longtable}
\usepackage{listings}
\usepackage{booktabs}
\usepackage{float}
\usepackage{makecell}
\usepackage{epstopdf}
\usepackage{pifont} 
\usepackage{wasysym}
\usepackage{tikz}
\usepackage{upgreek}
\usepackage[normalem]{ulem}
\usepackage{flafter}  
\usepackage{ulem}
\usepackage{color}
\usepackage{amsthm}
\usepackage[top=60pt,bottom=45pt,left=50pt,right=50pt]{geometry}
\usepackage{balance}
\newtheorem{thm}{Theorem}

\newtheorem{remark}{Remark}
\theoremstyle{plain}
\usepackage{lmodern,pdftexcmds,amsmath}
\usepackage[section,above,below]{placeins}
\usepackage{ifthen}
\newboolean{showcomments}
\setboolean{showcomments}{true}

\newcommand{\dennice}[1]{\ifthenelse{\boolean{showcomments}}
{\textcolor{blue}{Dennice says: #1}}{}}
\newcommand{\addcites}[0]{\ifthenelse{\boolean{showcomments}}
{\textcolor{blue}{(add cite(s))}}{}}
\newcommand{\chang}[1]{\ifthenelse{\boolean{showcomments}}
{\textcolor{red}{Chang says: #1}}{}}
\DeclareMathAlphabet{\mathsfbi}{OT1}{\sfdefault}{bx}{sl}
\DeclareMathVersion{sfletters}
\SetSymbolFont{letters}{sfletters}{OML}{ntxsfmi}{b}{it}

\newenvironment{myproof}{{\emph{Proof:}}}{\hfill$\square$}

\newtheorem{corollary}[thm]{Corollary} 

\newtheorem{example}{Example}

\makeatletter
\newcommand{\mathbfsbilow}[1]{%
  \text{\mathversion{sfletters}$\m@th#1$}%
}
\makeatother

\IEEEoverridecommandlockouts                              

\title{\LARGE \bf
Error bounds of constant gain least-mean-squares algorithms}

\author{Chang Liu$^{1,\dagger}$ and Antwan D. Clark$^{2}$ ~\IEEEmembership{Senior Member,~IEEE}
\thanks{$^{1}$ School of Mechanical, Aerospace, and Manufacturing Engineering, University of Connecticut, CT 06269, USA. $^{2}$ Department of Applied Mathematics and Statistics, Johns Hopkins University, MD 21218, USA.
        {\tt\small $\dagger$ chang\_liu@uconn.edu}}%
}

\begin{document}

\maketitle
\thispagestyle{empty}
\pagestyle{empty}

\begin{abstract}
Constant gain least-mean-squares (LMS) algorithms have a wide range of applications in trajectory tracking problems, but the formal convergence of LMS in mean square is not yet fully established. This work provides an upper bound on the constant gain that guarantees a bounded mean-squared error of LMS for a general design vector. These results highlight the role of the fourth-order moment of the design vector. Numerical examples demonstrate the applicability of this upper bound in setting a constant gain in LMS, while existing criteria may fail. We also provide the associated error bound, which can be applied to design vectors with linearly dependent elements. 
\end{abstract}

\begingroup
\allowdisplaybreaks

\section{Introduction}
\IEEEPARstart{L}{east} mean-squares (LMS) algorithms have a wide range of applications from signal processing and adaptive control \cite{uncini2015fundamentals} to geophysics \cite{robinson2000geophysical}. Particularly, understanding the impact of the constant gain in these algorithms is important in trajectory tracking and the estimation of temporal parameters where some applications include the works of  \cite{sun2009synchronization,liu2018time,liu2020hydrodynamics,zhang2019online,tsubota2023bifurcation} due to the fact that non-vanishing gains ensures the fidelity of adaptations to the time-dependent true parameter. Some advancements include the works of \cite{guo1995exponential, guo1995performance, guo1997necessary, ljung2012stochastic} where stochastic approximation theory has been used to find formal convergence criterion of recursive LMS algorithms in which the main requirement is that the indexed gain $a_k \rightarrow 0$ as $k \rightarrow \infty$ \cite{spall2005introduction}. Exploiting the concept of weak convergence \cite{kushner2003stochastic, kushner1981asymptotic}, researchers have found the connections between the constant gain and stochastic behavior \cite{macchi1983second,gerencser1995rate} as well as sensitivity to certain data properties \cite{wang2000lms}. Practical convergence benchmarks were also developed, which include the earlier works of \cite{widrow1985adaptive}, \cite{haykin1996adaptive}, and \cite{moon2000mathematical} who provided informal convergence arguments. Later, more  rules of thumb were proposed such as those related to mean-squared stability \cite{zhu2015error} and stochastic gradient methods with constant or non-decaying gains \cite{zhu2016tracking,zhu2018probabilistic,zhu2020stochastic,zhu2020error}. 

\par However, finding the bounds on these algorithms is still an open research problem. Although recent developments provided more measurable criterion, this is limited to sufficiently small gains and is not applicable for linearly dependent stochastic design vectors \cite{zhu2015error}. Here, we provide theoretical upper bounds on the constant gain of LMS algorithms that guarantees the bounded mean-squared error for a general design vector. Furthermore, these results provide upper bounds on the estimation errors as well as illustrate the correlations between the fourth-order moment of general design vectors. Our results, via synthetic and practical data, illustrate the promise of our proposed rule of thumb as well as demonstrate improvement over previous results for both general and linearly dependent stochastic design vectors \cite{widrow1985adaptive,zhu2015error}.

\par The rest of this paper is presented in the following manner. In Section \ref{sec:convergence}, we provide the upper bound on the gain to guarantee a bounded mean-squared error of LMS and the associated error bound. Section \ref{sec:numerical} provides examples demonstrating their fidelity. Section \ref{sec:conclusion} concludes this paper, where we highlight some future directions.  

\section{Error Bounds of LMS}
\label{sec:convergence}

\par Our analysis considers the simple linear regression model based on the scalar output measurement given by 

\begin{align}
    z_k=\boldsymbol{h}_k^T\boldsymbol{\theta}^*+\epsilon_k,\;k=1,2,...,n,
    \label{eq:measurement}
\end{align}

\noindent where $\boldsymbol{\theta}^*$ is the $m \times 1$ dynamic response parameter process of interest, $z_k$ is the scalar output measurement,  and $\epsilon_{k} = \mathcal{N} \left(0, \sigma_{\epsilon}^2 \right)$ is the sequence of independent normally distributed random noise. Furthermore, $\boldsymbol{h}_k$ is a $m\times 1$ stochastic design vector that is independent of $\epsilon_k$ consisting of a finite-magnitude covariance matrix $\mathbb{E}[\boldsymbol{h}_k\boldsymbol{h}_k^T] \succeq0$ ($\forall k\in \mathbb{N}$). The supposition in our analysis is that the covariance matrix is independent of $k$ and is represented by the Cholesky decompositions 

\begin{equation}
    \mathbb{E}[\boldsymbol{h}_k\boldsymbol{h}_k^T] = \boldsymbol{N}\boldsymbol{N}^T.
\label{CovarianceDecomp_k}
\end{equation}

\noindent Additionally, we consider the minimization of the loss function $\hat{L}_k=(z_k-\boldsymbol{h}_k^T\boldsymbol{\theta}^*)^2/2$ in which tracking the dynamic parameter is represented as 

\begin{align}
    \boldsymbol{\hat{\theta}}_{k}=&\boldsymbol{\hat{\theta}}_{k-1}-a\boldsymbol{h}_{k}(\boldsymbol{h}_{k}^T\boldsymbol{\hat{\theta}}_{k-1}-z_{k}),
    \label{eq:LMS}
\end{align}

\noindent where $a>0$ is the positive constant gain. Hence, we have the following results. 

\begin{thm}
If $\exists$ $\boldsymbol{P}\succ 0$ and $0<\chi<2/a$ such that:
\begin{align}
    a\mathbb{E}[\boldsymbol{h}_k\boldsymbol{h}_k^T\boldsymbol{P}\boldsymbol{h}_k\boldsymbol{h}_k^T]-\boldsymbol{P}\mathbb{E}[\boldsymbol{h}_k\boldsymbol{h}_k^T]-\mathbb{E}[\boldsymbol{h}_k\boldsymbol{h}_k^T]\boldsymbol{P}\preceq& -\chi \boldsymbol{P},
\label{eq:converge_req}
\end{align}
$\forall k\in \mathbb{N}^+$, then the LMS algorithm with the constant gain $a$ in equation (\ref{eq:LMS}) will provide estimation $\boldsymbol{\hat{\theta}}$ of true value $\boldsymbol{\theta}^*$ such that $\mathbb{E}[||{\boldsymbol{\hat{\theta}}}_k-\boldsymbol{\theta}^*||_2^2]\leq \Pi_k<\infty$ is bounded $\forall k\in \mathbb{N}^+$. Furthermore, the asymptotic limit of the upper bound is
\begin{align}
    \underset{k\rightarrow \infty}{\text{lim}}\Pi_k= \frac{a}{\chi \lambda_{\text{min}}(\boldsymbol{P})}\sigma_\epsilon^2\text{tr}\left[\boldsymbol{P} \mathbb{E}\big[ \boldsymbol{h}_{k}\boldsymbol{h}_k^T\big]\right].
    \label{eq:theorem_1_upper_bound_inf_t}
\end{align}
\label{thm:LMS_convergence}
\end{thm}

\begin{myproof}

Considering equations \eqref{eq:measurement} and \eqref{eq:LMS} the LMS estimation error $\boldsymbol{\tilde{\theta}}_k:=\boldsymbol{\hat{\theta}}_k-\boldsymbol{\theta}^*$ is given by 

\begin{align}
    \boldsymbol{\tilde{\theta}}_{k}=(\boldsymbol{I}-a\boldsymbol{h}_{k}\boldsymbol{h}_{k}^T)\boldsymbol{\tilde{\theta}}_{k-1}+a\boldsymbol{h}_{k}\epsilon_{k}, 
\label{eq:LMS_error}
\end{align}

\noindent $\boldsymbol{I}$ is the identity matrix. Since $\boldsymbol{P}\succ 0$ and noting that $V(\boldsymbol{\tilde{\theta}}_{k}) = \boldsymbol{\tilde{\theta}}_{k}^T\boldsymbol{P}\boldsymbol{\tilde{\theta}}_{k}\label{eq:V_theta_a}$ is a scalar quantity such that $\text{tr} \left[V\left(\boldsymbol{\tilde{\theta}}_{k}\right) \right] = \text{tr} \left[ \boldsymbol{\tilde{\theta}}_{k}^T\boldsymbol{P}\boldsymbol{\tilde{\theta}}_{k}\right] = V\left(\boldsymbol{\tilde{\theta}}_{k}\right)$ yields 

\begin{equation}
    V\left(\boldsymbol{\tilde{\theta}}_{k}\right) = \text{tr} \left[\boldsymbol{\tilde{\theta}}_{k}^T\boldsymbol{P}\boldsymbol{\tilde{\theta}}_{k}\right].
\label{VTheta:Def}
\end{equation}

\noindent Noting the right hand side of \eqref{VTheta:Def} with $\boldsymbol{\tilde{\theta}}_{k}$ defined by \eqref{eq:LMS_error} while also noting the following property

\begin{equation}
\text{tr} \left[\boldsymbol{\tilde{\theta}}_{k}^T\boldsymbol{P}\boldsymbol{\tilde{\theta}}_{k}\right] = \text{tr}\left[\boldsymbol{P}\boldsymbol{\tilde{\theta}}_{k}\boldsymbol{\tilde{\theta}}_{k}^T\right],
\label{eq:Trace_VT}
\end{equation}

\noindent the right hand side of \eqref{eq:Trace_VT} yields

\begin{align}
\text{tr} \left\{ \boldsymbol{P}\boldsymbol{\tilde{\theta}}_{k-1}\boldsymbol{\tilde{\theta}}_{k-1}^T  + \left(-a\boldsymbol{h}_{k}\boldsymbol{h}_{k}^T\right)\boldsymbol{P}\left(-a\boldsymbol{h}_{k}\boldsymbol{h}_{k}^T\right)\boldsymbol{\tilde{\theta}}_{k-1} \boldsymbol{\tilde{\theta}}_{k-1}^T \right. \nonumber \\
\left. +\boldsymbol{P}\left(-a\boldsymbol{h}_{k}\boldsymbol{h}_{k}^T\right)\boldsymbol{\tilde{\theta}}_{k-1} \boldsymbol{\tilde{\theta}}_{k-1}^T+ \left(-a\boldsymbol{h}_{k}\boldsymbol{h}_{k}^T\right)\boldsymbol{P}\boldsymbol{\tilde{\theta}}_{k-1}\boldsymbol{\tilde{\theta}}_{k-1}^T \right. \nonumber \\
\left. +a\epsilon_{k}\boldsymbol{P} \left(\boldsymbol{I}-a\boldsymbol{h}_{k}\boldsymbol{h}_{k}^T \right)\boldsymbol{\tilde{\theta}}_{k-1} \boldsymbol{h}_{k}^T \right. \nonumber \\
\left. +a\epsilon_{k}\boldsymbol{P}\boldsymbol{h}_{k}\boldsymbol{\tilde{\theta}}_{k-1}^T \left(\boldsymbol{I}-a\boldsymbol{h}_{k}\boldsymbol{h}_{k}^T \right)+a^2\epsilon_{k}^2\boldsymbol{P}\boldsymbol{h}_{k}\boldsymbol{h}_{k}^T\right\}.
\label{eq:VT_RHS}
\end{align}

\noindent Noting that $\boldsymbol{\tilde{\theta}}_{k}$ depends on the history $(\boldsymbol{H}_{k}, \boldsymbol{\Psi}_{k})$ where $\boldsymbol{H}_{k} =(\boldsymbol{h}_{k},\boldsymbol{h}_{k-1},...,\boldsymbol{h}_1)$, $\boldsymbol{\Psi}_{k} =(\epsilon_{k},\epsilon_{k-1},...,\epsilon_1)$ and that $\boldsymbol{h}_{k}$ and $\epsilon_{k}$ are measurable with respect to ($\boldsymbol{H}_{k}$, $\boldsymbol{\Psi}_{k}$), taking the conditional expectation of $V\left(\boldsymbol{\tilde{\theta}}_{k}\right)$ given ($\boldsymbol{H}_{k}$,$\boldsymbol{\Psi}_{k}$) while considering \eqref{eq:VT_RHS} yields 

\begin{subequations}
\begin{align}
    &\mathbb{E}\left[ V \left(\boldsymbol{\tilde{\theta}}_{k}\right)\Big|\boldsymbol{H}_{k},\boldsymbol{\Psi}_{k}\right] \nonumber \\
    =&\mathbb{E}\bigg\{\text{tr}\Big[\boldsymbol{P}\boldsymbol{\tilde{\theta}}_{k-1}\boldsymbol{\tilde{\theta}}_{k-1}^T+\left(-a\boldsymbol{h}_{k}\boldsymbol{h}_{k}^T \right)\boldsymbol{P} \left(-a\boldsymbol{h}_{k}\boldsymbol{h}_{k}^T \right)\boldsymbol{\tilde{\theta}}_{k-1} \boldsymbol{\tilde{\theta}}_{k-1}^T\nonumber\\
    &+\boldsymbol{P} \left(-a\boldsymbol{h}_{k}\boldsymbol{h}_{k}^T \right)\boldsymbol{\tilde{\theta}}_{k-1} \boldsymbol{\tilde{\theta}}_{k-1}^T+ \left(-a\boldsymbol{h}_{k}\boldsymbol{h}_{k}^T \right)\boldsymbol{P}\boldsymbol{\tilde{\theta}}_{k-1}\boldsymbol{\tilde{\theta}}_{k-1}^T\nonumber\\
    &+a\epsilon_{k}\boldsymbol{P} \left(\boldsymbol{I}-a\boldsymbol{h}_{k}\boldsymbol{h}_{k}^T \right) \boldsymbol{\tilde{\theta}}_{k-1} \boldsymbol{h}_{k} \nonumber\\
    &+a\epsilon_{k}\boldsymbol{P}\boldsymbol{h}_{k} \left(\boldsymbol{I}-a\boldsymbol{h}_{k}\boldsymbol{h}_{k}^T \right)\boldsymbol{\tilde{\theta}}_{k-1}\nonumber\\
    &+a^2\epsilon_{k}^2\boldsymbol{P}\boldsymbol{h}_{k}\boldsymbol{h}_{k}^T \Big]\Big|\boldsymbol{H}_{k},\boldsymbol{\Psi}_{k}\bigg\}\\ 
    =&\text{tr}\bigg\{\boldsymbol{P}\mathbb{E}\Big[\boldsymbol{\tilde{\theta}}_{k-1}\boldsymbol{\tilde{\theta}}_{k-1}^T \Big|\boldsymbol{H}_{k},\boldsymbol{\Psi}_{k}\Big] \nonumber\\
    &+a^2\boldsymbol{h}_{k}\boldsymbol{h}_{k}^T\boldsymbol{P}\boldsymbol{h}_{k}\boldsymbol{h}_{k}^T\mathbb{E}\left[\boldsymbol{\tilde{\theta}}_{k-1} \boldsymbol{\tilde{\theta}}_{k-1}^T \Big|\boldsymbol{H}_{k},\boldsymbol{\Psi}_{k}\right]\nonumber\\ 
    &-a\boldsymbol{P}\boldsymbol{h}_{k}\boldsymbol{h}_{k}^T\mathbb{E}\left[\boldsymbol{\tilde{\theta}}_{k-1} \boldsymbol{\tilde{\theta}}_{k-1}^T \Big|\boldsymbol{H}_{k},\boldsymbol{\Psi}_{k}\right]\nonumber\\
    &-a\boldsymbol{h}_{k}\boldsymbol{h}_{k}^T\boldsymbol{P}\mathbb{E}\left[\boldsymbol{\tilde{\theta}}_{k-1} \boldsymbol{\tilde{\theta}}_{k-1}^T\Big|\boldsymbol{H}_{k},\boldsymbol{\Psi}_{k}\right]\nonumber\\
    &+a\epsilon_{k}\boldsymbol{P}\left(\boldsymbol{I}-a\boldsymbol{h}_{k}\boldsymbol{h}_{k}^T\right)\mathbb{E}\left[\boldsymbol{\tilde{\theta}}_{k-1}\Big|\boldsymbol{H}_{k},\boldsymbol{\Psi}_{k}\right] \boldsymbol{h}_{k}^T\nonumber\\
    &+a\epsilon_{k}\boldsymbol{P}\boldsymbol{h}_{k}\mathbb{E}\left[\boldsymbol{\tilde{\theta}}_{k-1}^T\Big|\boldsymbol{H}_{k},\boldsymbol{\Psi}_{k}\right]\left(\boldsymbol{I}-a\boldsymbol{h}_{k}\boldsymbol{h}_{k}^T\right)\nonumber\\
    &+a^2\epsilon_{k}^2\boldsymbol{P}\boldsymbol{h}_{k}\boldsymbol{h}_{k}^T\bigg\}.
\label{eq:condition_expectation_c}
\end{align}
\end{subequations}

\noindent Applying iterative conditional expectation while also noting that $\epsilon_k = \mathcal{N} \left(0, \sigma_{\epsilon}^2\right)$ is independent and identically distributed (IID) gives

\begin{subequations}
\label{eq:E_V}
\begin{align}
    &\mathbb{E}\left[ V\left(\boldsymbol{\tilde{\theta}}_{k}\right)\right]\nonumber\\
    =&\text{tr}\bigg\{\Big(\boldsymbol{P}+a^2\mathbb{E}\left[\boldsymbol{h}_{k}\boldsymbol{h}_{k}^T\boldsymbol{P}\boldsymbol{h}_{k}\boldsymbol{h}_{k}^T\right]\nonumber\\
    &-a\boldsymbol{P}\mathbb{E}\left[\boldsymbol{h}_{k}\boldsymbol{h}_{k}^T\right]-a\mathbb{E}\left[\boldsymbol{h}_{k}\boldsymbol{h}_{k}^T\right]\boldsymbol{P}\Big)\mathbb{E}\left[\boldsymbol{\tilde{\theta}}_{k-1} \boldsymbol{\tilde{\theta}}_{k-1}^T\right]\nonumber \\
    &+0+0+a^2\sigma_\epsilon^2\boldsymbol{P}\mathbb{E}\left[\boldsymbol{h}_{k}\boldsymbol{h}_{k}^T\right]\bigg\}\\
    =&\text{tr}\left[\boldsymbol{P}\mathbb{E}\left[\boldsymbol{\tilde{\theta}}_{k-1} \boldsymbol{\tilde{\theta}}_{k-1}^T\right]\right]+\text{tr}\left[a\boldsymbol{Q}_{k}\mathbb{E}\left[\boldsymbol{\tilde{\theta}}_{k-1}\boldsymbol{\tilde{\theta}}_{k-1}^T\right]\right]\nonumber\\
    &+a^2\sigma_\epsilon^2\text{tr}\left[ \boldsymbol{P}\mathbb{E}\left[ \boldsymbol{h}_{k}\boldsymbol{h}_{k}^T\right]\right],
\end{align}
\end{subequations}

\noindent where $\boldsymbol{Q}_{k}= a\mathbb{E}[\boldsymbol{h}_{k}\boldsymbol{h}_{k}^T\boldsymbol{P}\boldsymbol{h}_{k}\boldsymbol{h}_{k}^T]-\boldsymbol{P}\mathbb{E}\left[\boldsymbol{h}_k\boldsymbol{h}_k^T\right]-\mathbb{E}\left[\boldsymbol{h}_k\boldsymbol{h}_k^T\right]\boldsymbol{P}$ is the left-hand side of \eqref{eq:converge_req}. Noting that $\mathbb{E}\big[\boldsymbol{\tilde{\theta}}_{k-1}\boldsymbol{\tilde{\theta}}_{k-1}^T\big]=\boldsymbol{M}\boldsymbol{M}^T$ by Cholesky decomposition while also considering the following trace property

\begin{align}
\text{tr}\left[a\boldsymbol{Q}_{k}\mathbb{E}\big[\boldsymbol{\tilde{\theta}}_{k-1}\boldsymbol{\tilde{\theta}}_{k-1}^T\big]\right] &= \text{tr}\left[a\boldsymbol{Q}_{k}\boldsymbol{M}\boldsymbol{M}^T\right] \nonumber \\
& = a\,\text{tr}\left[\boldsymbol{M}^T\boldsymbol{Q}_{k}\boldsymbol{M}\right].
\label{eq:Q_P_a1} 
\end{align}

\noindent Hence, employing condition \eqref{eq:converge_req}, equation \eqref{eq:Q_P_a1} becomes 

\begin{align}
\text{tr}\left[a\boldsymbol{Q}_{k}\mathbb{E}\left[\boldsymbol{\tilde{\theta}}_{k-1}\boldsymbol{\tilde{\theta}}_{k-1}^T\right]\right]
    \leq &-a\chi \,\text{tr}\left[\boldsymbol{M}^T\boldsymbol{P}\boldsymbol{M}\right] \nonumber \\
    =&-a\chi \,\text{tr}\left[\boldsymbol{P}\boldsymbol{M}\boldsymbol{M}^T\right] \nonumber \\
    =&-a\chi \,\text{tr}\left[\boldsymbol{P}\mathbb{E}\left[\boldsymbol{\tilde{\theta}}_{k-1}\boldsymbol{\tilde{\theta}}_{k-1}^T\right]\right]\nonumber\\
    =&-a\chi\mathbb{E}\left[ V(\boldsymbol{\tilde{\theta}}_{k-1})\right].
\label{eq:Q_P}
\end{align}

\noindent Combining equations \eqref{eq:E_V} and \eqref{eq:Q_P}, we have
\begin{subequations}
\label{eq:E_V_bound}
\begin{align}
&\mathbb{E}\left[ V(\boldsymbol{\tilde{\theta}}_{k})\right]\nonumber\\
 \leq&(1-a \chi)\mathbb{E}\left[ V(\boldsymbol{\tilde{\theta}}_{k-1})\right]+a^2\sigma_\epsilon^2\text{tr}\left[ \boldsymbol{P}\mathbb{E}\big[  \boldsymbol{h}_{k}\boldsymbol{h}_{k}^T\big]\right]\label{eq:E_V_bound_a}\\
 \leq& (1-a \chi)^{k}\mathbb{E}\left[ V(\boldsymbol{\tilde{\theta}}_{0})\right]+[1+(1-a\chi)+...(1-a\chi)^{k-1}]\nonumber\\
 &\times\,a^2\sigma_\epsilon^2\text{tr}\left[ \boldsymbol{P}\mathbb{E}\big[  \boldsymbol{h}_{k}\boldsymbol{h}_{k}^T\big]\right]\label{eq:E_V_bound_b}\\
 =& (1-a\chi)^{k}\mathbb{E}\left[ V(\boldsymbol{\tilde{\theta}}_{0})\right]+\frac{1-(1-a\chi)^{k}}{1-(1-a\chi)}\nonumber\\
 &\times\,a^2\sigma_{\epsilon}^2\text{tr}\left[ \boldsymbol{P}\mathbb{E}\big[  \boldsymbol{h}_{k}\boldsymbol{h}_{k}^T\big]\right],\label{eq:E_V_bound_c}
\end{align}
\end{subequations}

\noindent where the inequality \eqref{eq:E_V_bound_b} is obtained by applying inequality \eqref{eq:E_V_bound_a} iteratively and the inequality \eqref{eq:E_V_bound_c} is obtained via geometric series. Therefore, $\mathbb{E}\left[ V(\boldsymbol{\tilde{\theta}}_{k})\right]$ is bounded $\forall k\in \mathbb{N}^+$ as $|1-a\chi|<1$ based on {the fact that $0<\chi<2/a$}. Using the inequality $\lambda_{\text{min}}(\boldsymbol{P})\|\boldsymbol{\hat{\theta}}-\boldsymbol{\theta}^*\|_2^2\leq V(\boldsymbol{\tilde{\theta}}_k)$ as $\boldsymbol{P}\succ 0$, we have

\begin{align}
    \mathbb{E}\left[\|\boldsymbol{\hat{\theta}}_k-\boldsymbol{\theta}^*\|_2^2\right]\leq \mathbb{E}\left[ V(\boldsymbol{\tilde{\theta}}_{k})\right]/\lambda_{\text{min}}(\boldsymbol{P}),
    \label{eq:E_MSE_bound}
\end{align}

\noindent which leads to the bounded mean-squared error:
\begin{align}
\mathbb{E}\left[\|\boldsymbol{\hat{\theta}}_k-\boldsymbol{\theta}^*\|_2^2\right]&\leq\Pi_k:= \frac{(1-a\chi)^{k}}{\lambda_{\text{min}}(\boldsymbol{P})}\mathbb{E}\left[ V(\boldsymbol{\tilde{\theta}}_{0})\right]\nonumber\\
&+\frac{1-(1-a\chi)^{k}}{\chi \lambda_{\text{min}}(\boldsymbol{P})}a\sigma_{\epsilon}^2\text{tr}\left[ \boldsymbol{P}\mathbb{E}\big[  \boldsymbol{h}_{k}\boldsymbol{h}_{k}^T\big]\right].
\end{align}

Taking $ \underset{k\rightarrow \infty}{\text{lim}}$ $\Pi_k$, we have the asymptotic limit of the upper bound as
\begin{align}
     \underset{k\rightarrow \infty}{\text{lim}}\Pi_k=& \frac{a}{\chi \lambda_{\text{min}}(\boldsymbol{P})}\sigma_\epsilon^2\text{tr}\left[ \boldsymbol{P}\mathbb{E}\big[  \boldsymbol{h}_{k}\boldsymbol{h}_{k}^T\big]\right].
\end{align}
\end{myproof}

\noindent The case where $\boldsymbol{P} = \boldsymbol{I}$ produces the following corollary.

\begin{corollary}
If $\exists\, 0<\chi<2/a$ such that
\begin{align}
    a\mathbb{E}[\boldsymbol{h}_k\boldsymbol{h}_k^T\boldsymbol{h}_k\boldsymbol{h}_k^T]-2\mathbb{E}[\boldsymbol{h}_k\boldsymbol{h}_k^T]\preceq -\chi \boldsymbol{I} ,
    \label{eq:converge_req_identity}
\end{align}
then the LMS algorithm with the constant gain $a$ in equation (\ref{eq:LMS}) will provide estimation $\boldsymbol{\hat{\theta}}$ of true value $\boldsymbol{\theta}^*$ such that $\mathbb{E}[||{\boldsymbol{\hat{\theta}}}_k-\boldsymbol{\theta}^*||_2^2]\leq \Pi_k<\infty$ is bounded $\forall k\in \mathbb{N}^+$. Furthermore, the asymptotic limit of the upper bound is given by
\begin{align}
     \underset{k\rightarrow \infty}{\text{lim }}\Pi_k= \frac{a}{\chi}\sigma_\epsilon^2\text{tr}\left[ \mathbb{E}\big[ \boldsymbol{h}_{k}\boldsymbol{h}_k^T\big]\right].\label{eq:corollary_2_upper_bound_inf_t}
\end{align}
\label{cor:LMS_convergence_identity}
\end{corollary}

\noindent It is also important to note the following remarks. 

\begin{remark}
    If we consider additional assumptions, we have condition \eqref{eq:converge_req_identity} in Corollary \ref{cor:LMS_convergence_identity} the same as conditions reported in the literature as summarized in Table \ref{Table:AssumptionConvergence}.     \label{remark:additional_assumption}
\end{remark}

\begin{table}[htbp!]
	\centering
	\caption{}
	\label{Table:AssumptionConvergence}       
	\begin{tabular}{p{1.05cm}p{2.50cm}}
		\hline
		\multicolumn{1}{c}{\textbf{Assumption}} & \multicolumn{1}{c}{\textbf{Convergence Condition}}  \\
        \multicolumn{1}{c}{} & \multicolumn{1}{c}{\textbf{(with Reference)}}  \\
		\hline
            \multicolumn{1}{c}{} & \\
        \multicolumn{1}{l}{$\mathbb{E}\left[\boldsymbol{h}_k\boldsymbol{h}_k^T\boldsymbol{h}_k\boldsymbol{h}_k^T\right]$} & \multicolumn{1}{c}{} \\
		\multicolumn{1}{r}{$=\mathbb{E}\left[\boldsymbol{h}_k\boldsymbol{h}_k^T\right]\mathbb{E}\left[\boldsymbol{h}_k\boldsymbol{h}_k^T\right]$} & \multicolumn{1}{r}{$a<2 \Big /\left(\lambda_{\text{max}}\left\{\mathbb{E}\left[\boldsymbol{h}_k\boldsymbol{h}_k^T\right]\right\}\right)$} \\
    \\
        \multicolumn{1}{c}{} & \multicolumn{1}{c}{(\cite[p. 102]{widrow1985adaptive})}\\
		\hline
		\multicolumn{1}{c}{} & \\
        \multicolumn{1}{l}{$\text{tr} \left\{\mathbb{E}\left[\boldsymbol{h}_k\boldsymbol{h}_k^T\boldsymbol{h}_k\boldsymbol{h}_k^T\right]\right\}$} & \multicolumn{1}{r}{} \\
		\multicolumn{1}{r}{$=\Big(\text{tr}\left\{\mathbb{E}\left[\boldsymbol{h}_k\boldsymbol{h}_k^T\right]\right\}\Big)^2$} 	&  \multicolumn{1}{r}{$a<2 \Big /\text{tr}\left(\mathbb{E}\left[\boldsymbol{h}_k\boldsymbol{h}_k^T\right]\right)$} \\
  \\
   \multicolumn{1}{c}{} & \multicolumn{1}{c}{(\cite[p. 103]{widrow1985adaptive})}\\
		\hline
		\multicolumn{1}{c}{} & \\
        \multicolumn{1}{l}{$\mathbb{E}\left[\boldsymbol{h}_k\boldsymbol{h}_k^T\boldsymbol{h}_k\boldsymbol{h}_k^T\right]$} & \multicolumn{1}{l}{$a<\left(2 \lambda_{\text{min}} \, \left\{\mathbb{E}\left[\boldsymbol{h}_k\boldsymbol{h}_k^T\right]\right\}\right)$} \\
		\multicolumn{1}{r}{$ \preceq \Big(\lambda_{\text{max}}\left\{\mathbb{E}\left[\boldsymbol{h}_k\boldsymbol{h}_k^T\right]\right\}\Big)^2 \boldsymbol{I}$} &  \multicolumn{1}{r}{$ \div \left( \lambda_{\text{max}} \, \left\{\mathbb{E}\left[\boldsymbol{h}_k\boldsymbol{h}_k^T\right]\right\}\right)^{2}$} \\
  \\
   \multicolumn{1}{c}{} & \multicolumn{1}{c}{(\cite[Eq. (9)]{zhu2015error})}\\
		\hline
	\end{tabular}
\end{table}



\begin{remark}
\label{remark:zhu_spall_2015_4}
If the fourth-order moment term in \eqref{eq:converge_req_identity} in Corollary \ref{cor:LMS_convergence_identity} vanishes (i.e., $\mathbb{E}[\boldsymbol{h}_k\boldsymbol{h}_k^T\boldsymbol{h}_k\boldsymbol{h}_k^T]=\boldsymbol{0}$), then
\begin{align}
  \chi =2\lambda_{\text{min}}\left\{\mathbb{E}[\boldsymbol{h}_k\boldsymbol{h}_k^T]\right\}. 
\end{align}
Equation \eqref{eq:corollary_2_upper_bound_inf_t} in Corollary \ref{cor:LMS_convergence_identity} becomes
\begin{align}
     \underset{k\rightarrow \infty}{\text{lim }}\Pi_k= \frac{a\sigma_\epsilon^2\,\text{tr}\left\{\mathbb{E}[\boldsymbol{h}_k\boldsymbol{h}_k^T]\right\}}{2\lambda_{\text{min}}\left\{\mathbb{E}[\boldsymbol{h}_k\boldsymbol{h}_k^T]\right\}},
    \label{eq:zhu_spall_upper_bound_inf_t}
\end{align}
\noindent which is the same as the upper bound in \cite[equation (11)]{zhu2015error}. 
\end{remark}

\par The motivation of these results stem from analyzing the Lyapunov stability of dynamical systems \cite{boyd1994linear}. Specifically,  Theorem \ref{thm:LMS_convergence} provides sufficient conditions for convergence of the LMS mean-squared error, where $V(\boldsymbol{\tilde{\theta}}_{k})$ is an analogous Lyapunov function and the dynamical system is the dynamic behavior that is dependent on $k$. The case where $\boldsymbol{P}=\boldsymbol{I}$ (as presented in Corollary \ref{cor:LMS_convergence_identity}) provides not only a conservative estimate but also a connection to the energy stability \cite{liu2020input}. Furthermore, the remarks in this paper connect Theorem \ref{thm:LMS_convergence} and Corollary \ref{cor:LMS_convergence_identity} to existing results in \cite{widrow1985adaptive,zhu2015error} via considering the additional assumptions on $\boldsymbol{h}_k$. However, these assumptions are in general not guaranteed and thus the associated convergence conditions may fail when applied in practical applications.

\section{Numerical examples}
\label{sec:numerical}

In this section, we consider two numerical examples to obtain the upper bound of constant gain $a$ of LMS through applying Theorem \ref{thm:LMS_convergence} and Corollary \ref{cor:LMS_convergence_identity}, which will be compared with the upper bound in Remark \ref{remark:additional_assumption} according to \cite{widrow1985adaptive,zhu2015error}. Next, we will compare the asymptotic limit of the upper bound of mean-squared error $ \underset{k\rightarrow \infty}{\text{lim}}\Pi_k$ in Theorem \ref{thm:LMS_convergence} and Corollary \ref{cor:LMS_convergence_identity} against that in \cite{zhu2015error} as mentioned in Remark \ref{remark:zhu_spall_2015_4}. Although the synthetic Example \ref{ex:example1} is not considering a $\mathbb{E}[\boldsymbol{h}_k\boldsymbol{h}_k^T]$ varying over $k$, this example can still demonstrate the advantage of proposed Theorem \ref{thm:LMS_convergence} and Corollary \ref{cor:LMS_convergence_identity} compared with the literature \cite{widrow1985adaptive,zhu2015error}.

\begin{example}
We consider $\boldsymbol{h}_k\in \mathbb{R}^{2\times 1}$ as a two-dimensional random variables with zero mean and respective standard deviation $\sigma_1$ and $\sigma_2$, and the correlation coefficient $\rho$. Thus, we have covariance matrix
\begin{align}
    \mathbb{E}[\boldsymbol{h}_k\boldsymbol{h}_k^T]=\begin{bmatrix}
        \sigma_1^2 & \rho \sigma_1\sigma_2\\
        \rho\sigma_1\sigma_2 & \sigma_2^2
    \end{bmatrix}.
\end{align}
We denote $\boldsymbol{P}=\begin{bmatrix}
    p_{11} & p_{12}\\
    p_{21} & p_{22}
\end{bmatrix}$ leading to each component of $\mathbb{E}[\boldsymbol{h}_k\boldsymbol{h}_k^T \boldsymbol{P} \boldsymbol{h}_k\boldsymbol{h}_k^T]=:\begin{bmatrix}
    f_{11} & f_{12}\\
    f_{21} & f_{22}
\end{bmatrix}$ as
\begin{subequations}
\begin{align}
f_{11}=&3\sigma_1^4 p_{11}+k_2\sigma_1^2\sigma_2^2 p_{22}+k_3\sigma_1^3\sigma_2(p_{12}+p_{21}),\\
f_{12}=&\sigma_1^2\sigma_2^2 (p_{12}+p_{21})+k_3\sigma_1^3\sigma_2 p_{11}+k_3\sigma_1\sigma_2^3p_{22},\\
f_{21}=&k_2\sigma_1^2\sigma_2^2(p_{12}+p_{21})+k_3\sigma_1^3\sigma_2p_{11}+k_3\sigma_1 \sigma_2^3p_{22},\\
f_{22}=&3\sigma_2^4p_{22}+k_2\sigma_1^2\sigma_2^2p_{11}+k_3\sigma_1\sigma_2^3(p_{12}+p_{21}),
\end{align}
\end{subequations}
where $k_2=1+2\rho^2$ and $k_3=3\rho$. We consider $\epsilon_k\sim \mathcal{N}(0,\sigma_\epsilon^2)$ and fix $\sigma_\epsilon=0.1$. We consider four different sets of parameters $\sigma_1$, $\sigma_2$, and $\rho$ specified in Table \ref{tab:parameter_example1}. 

\begin{table}[htbp!]
    \centering
    \begin{tabular}{cccc}
    \hline
          & $\sigma_1$ & $\sigma_2$ & $\rho$ \\
          \hline
         Example \ref{ex:example1}A &1 & 1 &0 \\
         Example \ref{ex:example1}B & 1 & 2 & 0 \\
        Example \ref{ex:example1}C &  1 & 1 & 0.5 \\
         Example \ref{ex:example1}D &  1 & 1 & 1\\
         \hline
    \end{tabular}
    \caption{Parameters $\sigma_1$, $\sigma_2$ and $\rho$ of Examples \ref{ex:example1}A-\ref{ex:example1}D. }
    \label{tab:parameter_example1}
\end{table}

\label{ex:example1}
\end{example}

\begin{example}
We use the oboe reed data (\url{https://www.jhuapl.edu/ISSO/PDF-txt/reeddata-fit.prn}) as discussed in Section 3.4 and Exercise 3.16 of \cite{spall2005introduction}. Here, we consider the curvilinear model the same as equation (3.26) in \cite{spall2005introduction}. We use the sample mean to approximate $\mathbb{E}[\boldsymbol{h}_k\boldsymbol{h}_k^T]$ and $\mathbb{E}[\boldsymbol{h}_k\boldsymbol{h}_k^T\boldsymbol{h}_k\boldsymbol{h}_k^T]$ leading to
\begin{align}
    &\mathbb{E}[\boldsymbol{h}_k\boldsymbol{h}_k^T]\nonumber\\
    =&\begin{bmatrix}
     1.740 & 	3.850 &	2.347 & 	2.122\\
     3.850&	9.770&	5.206 &	4.910\\
     2.347&	5.206&	3.666&	2.936\\
     2.121&	4.910&	2.936&	3.585
    \end{bmatrix},\\
    &\mathbb{E}[\boldsymbol{h}_k\boldsymbol{h}_k^T\boldsymbol{h}_k\boldsymbol{h}_k^T]\nonumber\\
    =&\begin{bmatrix}
    35.274&	83.016&	51.333&	48.512\\
    83.016&	216.007&	120.071&	118.083\\
    51.333&	120.071&	83.068&	71.289\\
    48.512&	118.083&	71.289&	91.267
    \end{bmatrix}.
\end{align}
\label{ex:example4}
\end{example}

For each example, we apply Theorem \ref{thm:LMS_convergence} and Corollary \ref{cor:LMS_convergence_identity} to obtain the lowest upper bounds of the constant gain $\text{sup}\,a$ of LMS. For implementing equation (\ref{eq:converge_req})
in Theorem \ref{thm:LMS_convergence}, we employ YALMIP  \cite{lofberg2004yalmip} of version R20190425 with the semi-definite programming (SDP) solver Mosek of version 9.0 (\url{www.mosek.com}). The lowest upper bound of the constant gain $a$ is then obtained through a bisection search solving
\begin{align}
    \text{max}\;\; a, \;\;\text{s.t.}\;\; \text{Theorem \ref{thm:LMS_convergence} is satisfied}.
\end{align}
Similarly, the lowest upper bound of constant gain $a$ according to Corollary \ref{cor:LMS_convergence_identity} is obtained by solving 
\begin{align}
    \text{max}\;\; a, \;\;\text{s.t.}\;\; \text{Corollary \ref{cor:LMS_convergence_identity} is satisfied}.
\end{align}
This can be solved by the same toolbox as Theorem \ref{thm:LMS_convergence}, but Corollary \ref{cor:LMS_convergence_identity} can be also solved by directly computing eigenvalues without any semi-definite programming (SDP) solver. 

Table \ref{tab:upper_bound} compares the lowest upper bound of constant gain $\text{sup}\,a$ obtained through Theorem \ref{thm:LMS_convergence} and Corollary \ref{cor:LMS_convergence_identity} against informal arguments for convergence of the constant gain LMS \cite{widrow1985adaptive} as well as criteria in \cite{zhu2015error}. Here, the lowest upper bounds of the constant gain $\text{sup}\,a$ obtained from both Theorem \ref{thm:LMS_convergence} and Corollary \ref{cor:LMS_convergence_identity} are typically smaller than those obtained from informal arguments as $2\Big /\lambda_{\text{max}}\{\mathbb{E}[\boldsymbol{h}_k\boldsymbol{h}_k^T]\}$ and $2\Big /\text{tr}\{\mathbb{E}[\boldsymbol{h}_k\boldsymbol{h}_k^T]\}$ \cite{widrow1985adaptive}. This corresponds to the observation that $\left(\mathbb{E}[\boldsymbol{h}_k\boldsymbol{h}_k^T]\right)^2\neq\mathbb{E}[\boldsymbol{h}_k\boldsymbol{h}_k^T\boldsymbol{h}_k\boldsymbol{h}_k^T]$ in general, which provides a possible explanation for why the informal criteria \cite{widrow1985adaptive} typically fail to guide practical applications. Regarding Example \ref{ex:example1}A, the criteria $2\lambda_{\text{min}}\{\mathbb{E}[\boldsymbol{h}_k\boldsymbol{h}_k^T]\}\Big/\left(\lambda_{\text{max}}\{\mathbb{E}[\boldsymbol{h}_k\boldsymbol{h}_k^T]\}\right)^2$ \cite{zhu2015error} is the same as $2\Big /\lambda_{\text{max}}\{\mathbb{E}[\boldsymbol{h}_k\boldsymbol{h}_k^T]\}$ and thus also larger than those obtained by Theorem \ref{thm:LMS_convergence} and Corollary \ref{cor:LMS_convergence_identity}. Regarding Examples \ref{ex:example1}B-\ref{ex:example1}C, $2\lambda_{\text{min}}\{\mathbb{E}[\boldsymbol{h}_k\boldsymbol{h}_k^T]\}\Big /\left(\lambda_{\text{max}}\{\mathbb{E}[\boldsymbol{h}_k\boldsymbol{h}_k^T]\}\right)^2$ \cite{zhu2015error} gives the upper bound on $a$ in a similar order to Theorem \ref{thm:LMS_convergence} and Corollary \ref{cor:LMS_convergence_identity}. This $2\lambda_{\text{min}}\{\mathbb{E}[\boldsymbol{h}_k\boldsymbol{h}_k^T]\}\Big /\left(\lambda_{\text{max}}\{\mathbb{E}[\boldsymbol{h}_k\boldsymbol{h}_k^T]\}\right)^2$ is not applicable when $\mathbb{E}[\boldsymbol{h}_k\boldsymbol{h}_k^T]$ has a zero eigenvalue, which appears when some elements of $\boldsymbol{h}_k$ are linearly dependent \cite{zhu2015error} (see Example \ref{ex:example1}D).

\begin{widetext}
\begin{table}[htbp!]
    \centering
    \begin{tabular}{cccccc}
     \hline
     & Ex. \ref{ex:example1}A  & Ex. \ref{ex:example1}B &Ex. \ref{ex:example1}C & Ex. \ref{ex:example1}D &Ex. \ref{ex:example4}  \\
     \hline
     Theorem \ref{thm:LMS_convergence} & 0.5000 (C) & 0.1610 (C) & 0.4227 (C) &0.3333 (C) & 0.0875  \\
     Corollary \ref{cor:LMS_convergence_identity} & 0.5000 (C) & 0.1538 (C) & 0.4000 (C) &0.3333 (C) & 0.0716\\
      {$2\Big /\lambda_{\text{max}}\{\mathbb{E}[\boldsymbol{h}_k\boldsymbol{h}_k^T]\}$ } \cite{widrow1985adaptive} & 2.0000 (N) & 0.5000 (N) & 1.3333 (N) &1.0000 (N) & 0.1169\\
      {$2\Big /\text{tr}\{\mathbb{E}[\boldsymbol{h}_k\boldsymbol{h}_k^T]\}$ }\cite{widrow1985adaptive} & 1.0000 (N) & 0.4000 (N) & 1.0000 (N) & 1.0000 (N) & 0.1066\\
      {$2\lambda_{\text{min}}\{\mathbb{E}[\boldsymbol{h}_k\boldsymbol{h}_k^T]\}\Big /\left(\lambda_{\text{max}}\{\mathbb{E}[\boldsymbol{h}_k\boldsymbol{h}_k^T]\}\right)^2$ } \cite{zhu2015error} & 2.0000 (N) & 0.1250 (C) &0.4443 (C) & 0.0000 (-) & 0.0007 \\
      \hline
    \end{tabular}
    \caption{The lowest upper bound of constant gain $\text{sup}\,a$ of LMS obtained using Theorem \ref{thm:LMS_convergence}, Corollary \ref{cor:LMS_convergence_identity}, and arguments in \cite{widrow1985adaptive} and \cite{zhu2015error} for Examples \ref{ex:example1}A-\ref{ex:example1}D and Example \ref{ex:example4}, respectively. Inside of the brackets, `C' denotes a bounded terminal error $\mathbb{E}[\|\boldsymbol{\hat{\theta}}_k-\boldsymbol{\theta}^*\|_2^2]<10$, while `N' represents unbounded terminal error $\mathbb{E}[\|\boldsymbol{\hat{\theta}}_k-\boldsymbol{\theta}^*\|_2^2]>10^{8}$ for numerical Examples \ref{ex:example1}A-\ref{ex:example1}D after $k=10^4$ iteration using LMS with $a=\text{sup}\,a-10^{-4}$. `-' means LMS is not applicable due to zero gain. Here, $\mathbb{E}[\|\boldsymbol{\hat{\theta}}_k-\boldsymbol{\theta}^*\|_2^2]$ is averaged over $10^3$ replication.}
    \label{tab:upper_bound}
\end{table}
\end{widetext}

Table \ref{tab:upper_bound} shows that the upper bound of constant gain $a$ obtained from Corollary \ref{cor:LMS_convergence_identity} is smaller than that derived from Theorem \ref{thm:LMS_convergence}, but Corollary \ref{cor:LMS_convergence_identity} is much easier to solve, which can provide an initial guidance on practical applications. Here, we also note that the upper bounds obtained from Theorem \ref{thm:LMS_convergence} and Corollary \ref{cor:LMS_convergence_identity} are the same for Example \ref{ex:example1}, where the design vectors are uncorrelated random variables and having the same means and variances. In Example \ref{ex:example1}D with perfect correlated elements in the design vector, the upper bound of the constant gain $\text{sup}\, a$ obtained from Theorem \ref{thm:LMS_convergence} and Corollary \ref{cor:LMS_convergence_identity} also overlap. However, Theorem \ref{thm:LMS_convergence} has the potential to provide a less conservative upper bound on $a$ compared to Corollary \ref{cor:LMS_convergence_identity} (see Examples \ref{ex:example1}B and \ref{ex:example1}C).

Next, we run LMS using the constant gain $a=\text{sup}\,a-\xi$ with $\xi=10^{-4}$ according to a range of criteria in Table \ref{tab:upper_bound}. We set the true value of parameter $\boldsymbol{\theta}^*$ as $\begin{bmatrix}1 &1\end{bmatrix}^T$ and the initial estimation $\boldsymbol{\hat{\theta}}_0$ as a random vector $\sim \mathcal{N}(\boldsymbol{0},\boldsymbol{I}_2)$. Here, we run LMS with iteration $k=10^4$ and $10^3$ replications. The convergence based on the expectation of the terminal error $\mathbb{E}[\|\boldsymbol{\hat{\theta}}_k-\boldsymbol{\theta}^*\|_2^2]$ ($k=10^4$) is also indicated in the bracket of Table \ref{tab:upper_bound}. Here, we can see that the criteria $2\Big /\lambda_{\text{max}}\{\mathbb{E}[\boldsymbol{h}_k\boldsymbol{h}_k^T]\}$ and $2\Big /\text{tr}\{\mathbb{E}[\boldsymbol{h}_k\boldsymbol{h}_k^T]\}$ \cite{widrow1985adaptive} lead to the unbounded estimation error for Examples \ref{ex:example1}A-\ref{ex:example1}D, and the criteria $2\lambda_{\text{min}}\{\mathbb{E}[\boldsymbol{h}_k\boldsymbol{h}_k^T]\}\Big /\left(\lambda_{\text{max}}\{\mathbb{E}[\boldsymbol{h}_k\boldsymbol{h}_k^T]\}\right)^2$ \cite{zhu2015error} is also not sufficient to provide a guarantee of a bounded mean-squared error; see Example \ref{ex:example1}A. However, the upper bound on the constant gain $\text{sup}\,a$ obtained from Theorem \ref{thm:LMS_convergence} and Corollary \ref{cor:LMS_convergence_identity} are able to provide a bounded $\mathbb{E}[\|\boldsymbol{\hat{\theta}}_k-\boldsymbol{\theta}^*\|_2^2]$ for Examples \ref{ex:example1}A-\ref{ex:example1}D.

\begin{table}[htbp!]
    \centering
    \begin{tabular}{ccccc}
    \hline
    
         & Ex. \ref{ex:example1}A & Ex. \ref{ex:example1}B & Ex. \ref{ex:example1}C & Ex. \ref{ex:example1}D \\
         \hline
        Gain $a$ & 0.4999 & 0.1537 & 0.3999 & 0.3332  \\
        Eq. \eqref{eq:theorem_1_upper_bound_inf_t} in Thm. \ref{thm:LMS_convergence}& 25.0 & 0.049 &0.11 & 327 \\
       Eq.  \eqref{eq:corollary_2_upper_bound_inf_t} in Cor. \ref{cor:LMS_convergence_identity} & 25.0 & 1.01 & 10.66 & Inf \\
      Eq. \eqref{eq:zhu_spall_upper_bound_inf_t} & $0.0050$ &0.0038 & 0.0080 & Inf\\
        Simulation & 0.072 & 0.0071 &0.025 & 0.11 \\
       \hline
    \end{tabular}
    \caption{The selected gain $a=\text{sup}\,a-10^{-4}$ according to Corollary \ref{cor:LMS_convergence_identity} and associated asymptotic upper bound of $\mathbb{E}[\|\boldsymbol{\hat{\theta}}_k-\boldsymbol{\theta}^*\|_2^2]$ based on equation \eqref{eq:theorem_1_upper_bound_inf_t} in Theorem \ref{thm:LMS_convergence}, equation \eqref{eq:corollary_2_upper_bound_inf_t} in Corollary \ref{cor:LMS_convergence_identity} and equation \eqref{eq:zhu_spall_upper_bound_inf_t} based on \cite[equation (11)]{zhu2015error}. These are compared with $\mathbb{E}[\|\boldsymbol{\hat{\theta}}_k-\boldsymbol{\theta}^*\|_2^2]$ obtained from numerical simulations after $k=10^4$ iterations averaged over $10^3$ replications.}
    \label{tab:upper_bound_estimation_error}
\end{table}

We further analyze the asymptotic limit of the upper bound $\Pi_k$ as provided in equation \eqref{eq:theorem_1_upper_bound_inf_t} in Theorem \ref{thm:LMS_convergence}, equation \eqref{eq:corollary_2_upper_bound_inf_t} in Corollary \ref{cor:LMS_convergence_identity} and equation \eqref{eq:zhu_spall_upper_bound_inf_t} according to Ref. \cite{zhu2015error}. We firstly select a constant gain $a$ that is the upper bound according to Corollary \ref{cor:LMS_convergence_identity} minus $\xi=10^{-4}$ such that \eqref{eq:converge_req_identity} can be satisfied leading to a positive $\chi$ using Corollary \ref{cor:LMS_convergence_identity}. We then solve the following optimization problem 
\begin{align}
    \text{max}\;\;\chi,\;\; \text{s.t.}\;\; \text{Theorem \ref{thm:LMS_convergence} is satisfied}.
\end{align}
The resulting $\chi$ and $\boldsymbol{P}$ are then used to obtain the error bound in equation \eqref{eq:theorem_1_upper_bound_inf_t} of Theorem \ref{thm:LMS_convergence}. Similarly, we solve 
\begin{align}
    \text{max}\;\;\chi,\;\; \text{s.t.}\;\; \text{Corollary \ref{cor:LMS_convergence_identity} is satisfied},
\end{align}

\noindent where $\chi$ can be used to obtain error bound in \eqref{eq:corollary_2_upper_bound_inf_t} of Corollary \ref{cor:LMS_convergence_identity}. We also compute the error bound as \eqref{eq:zhu_spall_upper_bound_inf_t} according to \cite{zhu2015error}. The selected constant gain $a$ and resulting asymptotic upper bounds $  \underset{k\rightarrow \infty}{\text{lim}}\Pi_k$ are summarized in Table \ref{tab:upper_bound_estimation_error}. We also perform simulations and evaluate $\mathbb{E}\left[\|\boldsymbol{\hat{\theta}}_k-\boldsymbol{\theta}^*\|_2^2\right]$ at $k=10^4$ averaged over $10^3$ replications, which is reported also in Table \ref{tab:upper_bound_estimation_error}. Here, we can see that the upper bounds provided by Theorem \ref{thm:LMS_convergence} and Corollary \ref{cor:LMS_convergence_identity} are larger than the simulation results, while the upper bound according to \cite{zhu2015error} is violated in Examples \ref{ex:example1}A-\ref{ex:example1}C. In the work of \cite{zhu2015error} the fourth-order term is neglected assuming $a$ is small enough (see Remark \ref{remark:zhu_spall_2015_4}, which is likely to result in the failure of the error bound \cite{zhu2015error} as shown here). Moreover, Theorem \ref{thm:LMS_convergence} provides an error bound that is significantly tighter than that based on Corollary \ref{cor:LMS_convergence_identity} in Examples \ref{ex:example1}B-\ref{ex:example1}C. In Example \ref{ex:example1}D, the correlated elements of the design vector lead to $\lambda_{\text{min}}\{\mathbb{E}[\boldsymbol{h}_k\boldsymbol{h}_k^T]\}=0$ and thus equation \eqref{eq:zhu_spall_upper_bound_inf_t} leads to infinity upper bound, which is also mentioned as one limitation \cite{zhu2015error}. The upper bound in Corollary \ref{cor:LMS_convergence_identity} also similarly leads to an infinity upper bound. However, Theorem \ref{thm:LMS_convergence} is able to provide a finite upper bound, which benefited from a general matrix $\boldsymbol{P}$.

\begin{table}[htbp!]
    \centering
    \begin{tabular}{ccccc}
    \hline
     & $\theta_{\text{F}}$  & $\theta_{\text{AE}}$ & $\theta_{\text{TF}}$ & $\theta_{\text{TF}}$ \\
     \hline
     Theorem \ref{thm:LMS_convergence} & 0.330 & 0.086 & 0.012 & 0.010\\
     Corollary \ref{cor:LMS_convergence_identity} & 0.275 & 0.081 & 0.041 & 0.027\\
      {$2\Big/\lambda_{\text{max}}\{\mathbb{E}[\boldsymbol{h}_k\boldsymbol{h}_k^T]\}$  \cite{widrow1985adaptive}} & 302 & 3275 & 3049 & -3446\\
      {$2\Big /\text{tr}\{\mathbb{E}[\boldsymbol{h}_k\boldsymbol{h}_k^T]\}$   \cite{widrow1985adaptive}} & 533 & 426 & 278 & -193\\
      {$\frac{2\lambda_{\text{min}}\{\mathbb{E}[\boldsymbol{h}_k\boldsymbol{h}_k^T]\}}{(\lambda_{\text{max}}\{\mathbb{E}[\boldsymbol{h}_k\boldsymbol{h}_k^T]\})^2}$ } \cite{zhu2015error} & 0.082 & 0.184 & 0.112 & 0.106 \\
      Batch LS \cite{spall2005introduction} & 0.584 & 0.101 & 0.078 &0.034 \\
      Recursive LS \cite{spall2005introduction} &0.557 & 0.106 & 0.086  &0.036  \\
      \hline
    \end{tabular}
    \caption{Terminal estimation $\boldsymbol{\hat{\theta}}_{161}$ using constant gain $a=\text{sup}\,a-10^{-4}$ of LMS using $\text{sup}\,a$ obtained from Theorem \ref{thm:LMS_convergence}, Corollary \ref{cor:LMS_convergence_identity}, and informal arguments \cite{widrow1985adaptive} and criterion \cite{zhu2015error} for Example \ref{ex:example4} as reported in Table \ref{tab:upper_bound}. Results are compared with those obtained from batch least-squares (LS) and recursive LS using the same data set \cite{spall2005introduction}.}
    \label{tab:data}
\end{table}

For Example \ref{ex:example4} we further consider the data from practical measurements (\url{https://www.jhuapl.edu/ISSO/PDF-txt/reeddata-fit.prn}) as discussed in Section 3.4 and Exercise 3.16 in \cite{spall2005introduction}. The obtained upper bounds are also compared with those obtained from informal arguments \cite{widrow1985adaptive} as shown in Table \ref{tab:upper_bound}. In Table \ref{tab:data}, we report the terminal estimation $\boldsymbol{\hat{\theta}}_{\text{161}}$ of Example \ref{ex:example4} after using all available measurements. The initial estimation $\boldsymbol{\hat{\theta}}_0=\boldsymbol{0}$ is employed, which is the same as that used in Exercise 3.16 for LMS in \cite{spall2005introduction}. In Table \ref{tab:data}, we also compare the results obtained from the batch least-squares method and recursive least-squares as reported in  \cite[Table 3.2]{spall2005introduction}. Here, we can see that the upper bound of the constant gain proposed in this paper is also applicable in practical data, while the informal argument in \cite{widrow1985adaptive} is likely to have a large estimation error. The criterion $2\lambda_{\text{min}}\{\mathbb{E}[\boldsymbol{h}_k\boldsymbol{h}_k^T]\}\Big /(\lambda_{\text{max}}\{\mathbb{E}[\boldsymbol{h}_k\boldsymbol{h}_k^T]\})^2$ \cite{zhu2015error} leads to a relatively small upper bound of $a$ as shown in Table \ref{tab:upper_bound}, which may lead to a slow converging rate of LMS as shown in Table \ref{tab:data}. This suggests that Theorem \ref{thm:LMS_convergence} and Corollary \ref{cor:LMS_convergence_identity} are also applicable to practical data. 


\section{Conclusion and future work}
\label{sec:conclusion}

Theorem \ref{thm:LMS_convergence} and Corollary \ref{cor:LMS_convergence_identity} propose upper bounds on the constant gain of LMS that guarantees a bounded mean-squared error. These sufficient conditions highlight the role of fourth-order moments in the design vector. We demonstrate the applicability of these upper bounds numerically based on synthetic and practical data, while existing criteria \cite{widrow1985adaptive,zhu2015error} introduce additional assumptions on the design vector and may lead to unbounded estimation results within constant gain LMS. The main result in Theorem \ref{thm:LMS_convergence} is computed based on semi-definite programming (SDP), while Corollary \ref{cor:LMS_convergence_identity} does not require an SDP solver which is able to provide initial guidance of the upper bound on the gain. However, Theorem \ref{thm:LMS_convergence} has the potential to provide a tighter upper bound on constant gain $a$ than Corollary \ref{cor:LMS_convergence_identity} which may lead to a faster convergence rate. 

These theoretical results also provide upper bounds of estimation error that demonstrate consistency with numerical examples, which is an improvement of \cite{zhu2015error}.  Additionally, the error bound of estimation error in Theorem \ref{thm:LMS_convergence} is applicable when elements of the design vector are linearly dependent, while both Corollary \ref{cor:LMS_convergence_identity} and Ref. \cite{zhu2015error} fail to provide a finite upper bound (Example \ref{ex:example1}D). Moreover, the numerical examples further demonstrate that 
Theorem \ref{thm:LMS_convergence} and Corollary \ref{cor:LMS_convergence_identity} provide the same lowest upper bound on the constant gain and estimation error when elements of the design vector are independent with the same variance (Example \ref{ex:example1}A).

This work can be naturally extended to consider the inherent dependence between $\{\boldsymbol{h}_k\}$ and $\{\epsilon_k\}$, where the associated fourth-order moment $\mathbb{E}[\epsilon_k^2\boldsymbol{h}_k\boldsymbol{h}_k^T]$ is also likely to be important; see the last term in \eqref{eq:condition_expectation_c}. Another avenue of future work is to further obtain the error bounds with a time-varying $\boldsymbol{\theta}^*$ and covariance matrix $\mathbb{E}[\boldsymbol{h}_k\boldsymbol{h}_k^T]$ \cite{zhu2015error,zhu2020error}, which is important in trajectory tracking applications. Convergence rate is another direction of future analysis.  

\section*{Acknowledgement}
C.L. would like to acknowledge the support from UConn Quantum Innovation Seed Grants and NASA Connecticut Space Grant Consortium Faculty Research Award P-2104 during the completion of this work. 

\balance
\bibliography{main}
\bibliographystyle{IEEEtran}
\end{document}